\documentclass[preprint,prd,amsmath,amssymb,nofootinbib,tightenlines,floatfix,letterpaper,superscriptaddress]{revtex4-1}
\usepackage[dvips]{graphics}
\usepackage{epsfig}
\usepackage{latexsym}
\usepackage{slashed}
\usepackage{color}
\newcommand{\be}{\begin{equation}}
\newcommand{\ee}{\end{equation}}
\newcommand{\bea}{\begin{eqnarray}}
\newcommand{\eea}{\end{eqnarray}}

\begin{document}
\preprint{NUHEP-TH/12-05}

\title{Dark Matter From Weak Polyplets}

\author{Andr\'{e} de Gouv\^{e}a\footnote{Electronic address:  degouvea@northwestern.edu}
}
\affiliation{Department of Physics and Astronomy, Northwestern University, Evanston, IL 60208 USA}

\author{Wei-Chih Huang\footnote{Electronic address:  whuang@sissa.it}
}
\affiliation{SISSA and INFN-sezione di Trieste, Via Bonomea 265,
34136 Trieste, Italy}

\author{Jennifer Kile\footnote{
	Electronic address: jenkile@northwestern.edu}
}
\affiliation{Department of Physics and Astronomy, Northwestern University, Evanston, IL 60208 USA}

\begin{abstract}
The addition of new multiplets of fermions charged under the Standard Model gauge group is investigated, with the aim of identifying a possible dark matter candidate.  These fermions are charged under $SU(2)\times U(1)$, and their quantum numbers are determined by requiring all new particles to obtain masses via Yukawa couplings and all triangle anomalies to cancel as in the Standard Model; more than one multiplet is required and we refer to such a set of these multiplets as a {\it polyplet}.  For sufficiently large multiplets, the stability of the dark matter candidate is ensured by an accidental symmetry; for clarity, however, we introduce a model with a particularly simple polyplet structure and stabilize the dark matter by imposing a new discrete symmetry.  We then explore the features of this model; constraints from colliders, electroweak precision measurements, the dark matter relic density, and direct detection experiments are considered.  We find that the model can accommodate a viable dark matter candidate for large Higgs boson masses; for $m_H\sim 125$ GeV, a subdominant contribution to the dark matter relic density can be achieved.
\end{abstract}

\maketitle

\section{Introduction}
\label{sec:intro}
The Standard Model (SM) of particle physics has demonstrated astounding success in describing phenomena at energy scales below a few hundred GeV.  However, despite this success, there are significant questions which the SM does not address adequately.  Some of these, such as the hierarchy problem and the origin of flavor, arise as theoretical issues within the SM, while others, such as the tiny but nonzero value of the cosmological constant, the origin of neutrino masses, and the identity of dark matter (DM), reflect discrepancies between observation and the SM prediction.  All of these questions indicate that, despite the remarkable success of the SM, it is not a complete theory of nature.

Given both the successes and the shortcomings of the SM, it makes sense to consider models of new physics which address one or more of the above questions while not radically altering the structure of the SM.  Following this philosophy, we investigate the possibility of adding new fermions, including a possible DM candidate, to the SM without altering its gauge group, $SU(3)\times SU(2)\times U(1)$.  (For examples of works where exotic fermion multiplets have been added to the SM, see \cite{Knochel:2011ng,Foot:1988qx,Foot:1988te,Foot:1988qu}.)  While all fermions in the SM transform as singlets or as the fundamental representations under the $SU(3)$ color and the $SU(2)$ weak interactions, higher-dimensional representations are possible.  Here, we investigate the possibility of new fermions which transform as higher-dimensional multiplets under $SU(2)$.  We will, in fact, add more than one new multiplet to the SM, and we will refer to this set of multiplets as a ``polyplet''.

There are several advantages to adding higher-dimensional $SU(2)$ fermion multiplets to the SM.  The first is that we can hope to arrange the quantum numbers of the multiplets such that they obtain their masses only via Yukawa couplings with the SM Higgs boson; thus our model would contain no new mass scales and would predict all of its new particles to be near the electroweak scale.  Second, if the dimensionality of these new multiplets is large, with isospin greater than $1$, it is impossible to write down dimension-four mass terms which mix these new fermions with those of the SM.  In addition to simplifying the model, this implies that the lightest member of the polyplet cannot decay to SM particles through renormalizable operators; this is useful as we hope that our polyplet will contain a DM candidate.\footnote{We note, however, that, for simplicity, we will work with a polyplet whose largest isospin is $1$ and introduce a new conserved quantum number to disallow the unwanted operators.}  For related works in which $SU(2)$ multiplets have been added to the SM in order to solve the DM puzzle, see \cite{Cirelli:2005uq,Hill:2011be,Chen:2011bc,Garny:2011ii,Kajiyama:2011gu}.

Several considerations must be taken into account when introducing new $SU(2)$ multiplets to the SM.  First, constraints on the existence of new neutral or charged massless particles from colliders \cite{pdg:2012} and cosmology \cite{Cyburt:2004yc,Komatsu:2010fb} imply that all members of these new multiplets must be given masses.  We choose the quantum numbers of these new multiplets such that all masses are acquired via Yukawa couplings with the SM Higgs; we wish to avoid dimension-3 mass terms and we disallow mixing with the SM fermions.  Additionally, we wish to avoid the possibility of a charged stable particle (see \cite{Langacker:2011db,SanchezSalcedo:2010ev,Berger:2008ti} and references therein for constraints on charged relics), so we insist that the lightest new particle is electrically neutral.   Finally, as we are interested in possible DM candidates, we only consider colorless particles.  

After giving masses to our new fermions, we explore their phenomenology.  Because their transformations under the SM gauge group are specified, their interactions are largely predetermined, although some couplings are tunable via mass mixing.  We consider present experimental bounds on these new higher-dimensional multiplets and find that they are allowed for a wide range of new particle masses.   We also find, for large Higgs boson masses, that our DM candidate can achieve the observed relic density via annihilation to gauge bosons and top quarks while simultaneously evading direct detection constraints; for $m_H\sim 125$ GeV, our DM candidate can only comprise a subdominant but non-negligible (${\cal O}(10\%)$) component of the observed relic density.

We organize the remainder of this paper as follows.  In Section \ref{sec:model}, we introduce our model, and discuss its particle content, couplings, and masses, and restrictions put on the model by requiring anomaly cancellation.  We then consider constraints from searches at colliders and precision electroweak measurements in Section \ref{sec:const}.  In Section \ref{sec:dm}, we investigate constraints placed by DM direct search experiments, and, finally, in Section \ref{sec:conc}, we conclude.

\section{The Model}
\label{sec:model}
\subsection{Notation and Basic Considerations}
We begin this section by specifying our notation and discussing some basic considerations which must be taken into account in order to make all of the particles in our model massive.  The polyplet we will consider contains four fermion multiplets, charged under $SU(2)\times U(1)$, which we will denote as left-handed chiral spinors $\chi_{n_i,Y_i}$. Here we label our multiplets with their number of components $n_i$ (where $n_i=2I_i+1$, with $I_i$ the multiplet isospin) and hypercharge $Y_i$. 

In order to evade the constraints on the existence of new massless particles, we must arrange masses for all of the particles in our model.  At the same time, we wish for our model to be predictive; therefore, we insist that all fermion mass terms be dimension-4 Yukawa couplings to the (hypercharge $1/2$) SM Higgs doublet $H$
\begin{eqnarray}
\label{eq:yuk}
\chi_{n_i,Y_i} &\times& \chi_{n_j,-Y_i-\frac{1}{2}} \times H_{\frac{1}{2}}, \nonumber\\
\chi_{n_i,Y_i} &\times& \chi_{n_j,-Y_i+\frac{1}{2}} \times H^{*}_{-\frac{1}{2}},
\end{eqnarray}
where $\times$ is the multiplication needed to render the mass terms $SU(2)$-invariant and the value of $Y_j$ has been determined in terms of $Y_i$ by hypercharge conservation.  After electroweak symmetry breaking, 
\begin{equation}
H\rightarrow \frac{1}{\sqrt{2}} \left(\begin{array}{c} 0 \\ v \end{array}\right),
\end{equation}
where $v\sim 246$ GeV; we thus expect the fermion masses to be tied to the electroweak scale.  

Because the SM Higgs field is an isospin doublet, two $\chi$ multiplets can couple to each other via a Yukawa term only if their isospins differ by $\pm 1/2$ or, equivalently, if $n_j=n_i\pm 1$ in Eq. (\ref{eq:yuk}).  Thus, $\chi$ multiplets with half-integer isospin (or even $n_i$) will only be coupled to those with integer isospin (or odd $n_i$), and vice-versa.  This implies that, in order for all particles in these multiplets to obtain masses, we must have an equal number of particles in even-$n_i$ multiplets and in odd-$n_i$ multiplets.  

Let us consider the possible Yukawa couplings of a multiplet $\chi_{n,Y}$ with $n$ components.   Let us assume it couples to another multiplet $\chi_{n-1,Y'}$:
\begin{equation}
\chi_{n,Y} \times \chi_{n-1,Y'} \times H^{(*)}_{\pm\frac{1}{2}}.
\end{equation} 
However, as $\chi_{n,Y}$ contains one more component than $\chi_{n-1,Y'}$, this cannot render every component of $\chi_{n,Y}$ massive.  Thus, $\chi_{n,Y}$ must couple to another multiplet.  We can either couple $\chi_{n,Y}$ to another multiplet with $n-1$ components, or to one with $n+1$ components.  If we couple it to another multiplet with $n-1$ components, our fourth multiplet must contain $n-2$ components in order to balance the number of particles in even-$n$ and odd-$n$ multiplets, and our model contains multiplets of size $n$, $n-1$, $n-1$ and $n-2$.  If we instead couple $\chi_{n,Y}$ to a multiplet with $n+1$ components, we obtain multiplets of size $n$, $n$, $n-1$, and $n+1$.  These two choices are clearly degenerate by a redefinition of $n$.  We therefore can write the four members of our polyplet as:
\begin{equation}
\chi_{n,Y}, \chi_{n,Y'}, \chi_{n+1,Y^+}, \chi_{n-1,Y^-}.
\end{equation}

Next, we note that, for a given value of $Y^+$, the members of $\chi_{n+1,Y^+}$ will have $n+1$ distinct values for their electric charge, $Q=I_{(3)}+Y$.  In order for all of these fields to become massive via Yukawa couplings with $\chi_{n,Y}$ and $\chi_{n,Y'}$, these same values of $Q$, but with opposite sign, must be represented in the components of $\chi_{n,Y}$ and $\chi_{n,Y'}$, which requires that $Y\ne Y'$.  We therefore take the Yukawa couplings of $\chi_{n+1,Y^+}$ as
\begin{eqnarray}
\label{eq:yukplus}
&&\chi_{n,Y} \times \chi_{n+1,Y^+} \times H^{*}_{-\frac{1}{2}}, \nonumber\\
&&\chi_{n,Y'} \times \chi_{n+1,Y^+} \times H_{\frac{1}{2}},
\end{eqnarray}
from which we can deduce that 
\begin{eqnarray}
Y + Y^+ -\frac{1}{2} = 0,\nonumber\\
Y' + Y^+ +\frac{1}{2} = 0,\\ 
Y'=Y-1.\nonumber
\end{eqnarray}

Now we consider the Yukawa couplings involving $\chi_{n-1,Y^-}$.  There are three possibilities for these couplings.  First, $\chi_{n-1,Y^-}$ could couple to both $\chi_{n,Y}$ and $\chi_{n,Y'}$.  In this case, we yield couplings analogous to those in Eq. (\ref{eq:yukplus}), and we trivially obtain $Y^-=Y^+$.\footnote{It is also possible to include only one of these two couplings, but this will not change the resulting relationships between the multiplet hypercharges.}  However, it is also possible that $\chi_{n-1,Y^-}$ couples only to $\chi_{n,Y}$ or only to $\chi_{n,Y'}$.  In these cases, one can have either a coupling of the form
\begin{equation}
\chi_{n,Y} \times \chi_{n-1,Y^-} \times H_{\frac{1}{2}},
\end{equation}
or one of the form
\begin{equation}
\chi_{n,Y'} \times \chi_{n-1,Y^-} \times H^{*}_{-\frac{1}{2}},
\end{equation}
which yields $Y^-=Y^+-1$ or $Y^-=Y^++1$, respectively.  We will thus adopt the following notation for our fermion multiplets:
\begin{equation}
\label{eq:qnums}
\chi_{n,Y}, \chi_{n,Y-1}, \chi_{n+1,\frac{1}{2}-Y}, \chi_{n-1,\frac{1}{2}-Y+A},
\end{equation}
where $A=0, 1$ or $-1$.

With these preliminaries out of the way, we will now explore the constraints on $n$, and the multiplet hypercharges imposed by anomaly cancellation.

\subsection{Anomaly Cancellation}
When building models which contain new particles or interactions, one must take care not to ruin the cancellation of triangle anomalies which occurs in the SM.  (For a related work on anomaly-free sets of fermions, see \cite{Batra:2005rh}.)  As our multiplets are not charged under $SU(3)$, there are only three anomaly relations relevant here:  $U(1)$-gravity-gravity, $U(1)^3$, and $U(1)\times SU(2)^2$.  These three relations take the form
\begin{eqnarray}
&&\sum_{\mbox{multiplets  }i}  n_i Y_i=0, \nonumber\\
&&\sum_{\mbox{multiplets  }i}  n_i Y_i^3=0, \\
&&\sum_{\mbox{multiplets  }i}  C(i) Y_i=0,\nonumber
\label{eq:anoms}
\end{eqnarray}
where 
$C(i)=\frac{1}{3}I_i(I_i+1)(2I_i+1)$. 

When we apply these constraints to the quantum numbers of our multiplets shown in Eq. (\ref{eq:qnums}), we obtain
\begin{eqnarray}
\label{eq:anomy}
&&  (n-1) A=0 \nonumber,\\
&&  3(n-1)A (\frac{1}{2}-Y)^2+3(-\frac{n}{2}+A^2 (n-1)) (\frac{1}{2}-Y)+(n-1)A^3=0,\\   
&& ( f(n+1) + f(n-1) -2f(n))(\frac{1}{2}-Y)+f(n-1)A=0,\nonumber
\end{eqnarray}
where, for clarity, we have written the third relation in terms of $f(x)=(x-1) x (x+1)/12 $.  Note that $f(n+1) + f(n-1) -2f(n)=n/2$.

Let us examine the first relation in Eqs. (\ref{eq:anomy}).  This equation has two solutions; $n=1$ or $A=0$.  Plugging either of these solutions into the other two equations in (\ref{eq:anomy}), we obtain $Y=1/2$.  However, we note that, for $Y=1/2$, odd $n$ implies that our multiplets contain particles of half-integer charge.  As such fractionally-charged (and stable) particles are observationally ruled out \cite{Langacker:2011db,SanchezSalcedo:2010ev,Berger:2008ti}, we insist on even $n$ and, therefore, $A=0$, yielding the multiplets\footnote{We note that for even $n$, $\mbox{Tr } I_{(3)}^2$, when summed over all multiplets, yields an integer; thus, this choice also satisfies the Witten anomaly condition \cite{Witten:1982fp}.}
\begin{equation}
\chi_{n,+\frac{1}{2}}, \chi_{n,-\frac{1}{2}}, \chi_{n+1,0}, \chi_{n-1,0}.
\end{equation}

This set of quantum numbers, however, allows dimension-3 mass terms like $m \chi_{n-1,0} \times \chi_{n-1,0}$, $m \chi_{n+1,0} \times \chi_{n+1,0}$, and $m \chi_{n,+\frac{1}{2}} \times \chi_{n,-\frac{1}{2}}$, where again $\times$ is the multiplication needed to make an $SU(2)$ singlet.   We would like to disallow these terms, in order to have a predictive model with masses set by the electroweak scale.  We can do this either by introducing additional fields to loosen the constraints set by anomaly cancellation, or by assigning a new conserved charge to our multiplets.  For simplicity, here, we will impose the conservation of a new charge, which we refer to as ``pseudo-lepton number''.  A discussion of how we can instead loosen the restrictions imposed by anomaly cancellation via the introduction of additional fields is given in the Appendix.    

With all of the hypercharges now specified, we briefly discuss whether or not the four allowed Yukawa couplings are sufficient to render every member of the polyplet massive.  To see this, we revisit the couplings of $\chi_{n+1,Y^+}$ given in Eq. (\ref{eq:yukplus}), which now take the form
\begin{equation}
\label{eq:yukplussp2}
y \mbox{ }\chi_{n,-\frac{1}{2}} \times \chi_{n+1,0} \times H_{\frac{1}{2}},
\end{equation}
and
\begin{equation}
\label{eq:yukplussp1}
y' \mbox{ }\chi_{n,\frac{1}{2}} \times \chi_{n+1,0} \times H^{*}_{-\frac{1}{2}},
\end{equation}
where we have introduced two Yukawa couplings $y$ and $y'$.  The couplings in Eq. (\ref{eq:yukplussp2}) will couple the components in $\chi_{n,-\frac{1}{2}}$ to the first $n$ components of $\chi_{n+1,0}$,
\begin{equation}
C_1\chi_{n,-\frac{1}{2}}(n) \chi_{n+1,0}(1) + C_2\chi_{n,-\frac{1}{2}}(n-1) \chi_{n+1,0}(2)....+ C_{n}\chi_{n,-\frac{1}{2}}(1) \chi_{n+1,0}(n),
\end{equation}
while those in Eq. (\ref{eq:yukplussp1}) will couple the components of $\chi_{n,\frac{1}{2}}$ the last $n$ components of $\chi_{n+1,0}$,
\begin{equation}
C'_2\chi_{n,\frac{1}{2}}(n) \chi_{n+1,0}(2) + C'_3\chi_{n,\frac{1}{2}}(n-1) \chi_{n+1,0}(3)....+ C'_{n+1}\chi_{n,\frac{1}{2}}(1) \chi_{n+1,0}(n+1),
\end{equation}
where the coefficients $C_i$ and $C_i'$ are products of $v/\sqrt{2}$, Clebsch-Gordon coefficients, and either $y$ or $y'$, respectively.  Two members of $\chi_{n+1,0}$, $\chi_{n+1,0}(1)$ and $\chi_{n+1,0}(n+1)$, couple to only one component of either $\chi_{n,\frac{1}{2}}$ or $\chi_{n,-\frac{1}{2}}$ and are thus mass eigenstates.  However, for $1<i<n+1$, $\chi_{n+1,0}(i)$ couples to a linear combination of one component from each $n$-plet,
\begin{equation}
\left( C_i  \chi_{n,-\frac{1}{2}}(n+1-i) + C_i' \chi_{n,\frac{1}{2}}(n+2-i) \right)\chi_{n+1,0}(i).
\end{equation}
while the orthogonal combination of $\chi_{n,-\frac{1}{2}}(n+1-i)$ and $\chi_{n,\frac{1}{2}}(n+2-i)$ remains massless.  When we include the analogous couplings of $\chi_{n-1,0}$,
\begin{equation}
\left( D_i  \chi_{n,-\frac{1}{2}}(n+1-i) + D_i' \chi_{n,\frac{1}{2}}(n+2-i) \right)\chi_{n-1,0}(i-1),
\end{equation}
these previously massless combinations will receive Yukawa couplings as long as $D'_i/D_i\neq C_i'/C_i$.  As the coefficients $C_i$, $C_i'$, $D_i$, $D_i'$ are functions of arbitrary yukawa couplings, it is reasonable to expect that they can generally be adjusted to render all of the fields massive, as long as none of the relevant Clebsch-Gordon coefficients are zero.
 
Up until this point, we have not specified $n$.  We will take the simplest polyplet which yields particles with integer electric charge, $n=2$.  We assign $\chi_{2,+\frac{1}{2}}$ and $\chi_{2,-\frac{1}{2}}$ a pseudo-lepton number of $1$ and $\chi_{3,0}$ and $\chi_{1,0}$ a pseudo-lepton number of $-1$.  All SM particles have a pseudo-lepton number of $0$.  We note that these assignments also prevent our multiplets from mixing with the SM leptons. 

As a side note, we briefly mention that the quantum numbers for the SM quarks satisfy the relations given in Eq. (\ref{eq:qnums}) for $n=1$, $Y=1/3$; here, $\chi_{1,\frac{1}{3}}$, $\chi_{1,-\frac{2}{3}}$ and $\chi_{2,\frac{1}{6}}$ correspond to $d_R^c$, $u_R^c$, and the $SU(2)$ quark doublet, respectively.\footnote{Note that for $n=1$, there is no fourth multiplet, as it would have $n-1=0$ components.}  Additionally, the SM leptons form an incomplete $n=1$, $Y=1$ polyplet, with $\chi_{1,1}$ corresponding to $e^c_R$, and $\chi_{2,-\frac{1}{2}}$ to the lepton doublet; the addition of $\chi_{1,0}$ (i.e., a right-handed neutrino), would give the neutrino a Dirac mass.  Of course, in the SM, anomaly cancellation is achieved by having contributions from the quark and lepton sectors cancel against each other, unlike in our case here.

We now move on to discuss some of the phenomenology of the $n=2$ model introduced above.

\subsection{Features of the model}

After the dust has settled, our model contains $8$ chiral states.  $\chi_{2,+\frac{1}{2}}$ and $\chi_{2,-\frac{1}{2}}$ each contain one charged and one neutral state,
\begin{eqnarray}
\chi_{2,+\frac{1}{2}} &=& \left(\begin{array}{c} \chi_{2,+\frac{1}{2}}^+ \\ \chi_{2,+\frac{1}{2}}^0  \end{array} \right),\nonumber\\
\chi_{2,-\frac{1}{2}} &=& \left(\begin{array}{c} \chi_{2,-\frac{1}{2}}^0 \\ \chi_{2,-\frac{1}{2}}^{-}  \end{array} \right),
\end{eqnarray}
where we have introduced superscripts to indicate the electric charge of each member of the multiplet.  Being $SU(2)$ doublets, $\chi_{2,+\frac{1}{2}}$ and $\chi_{2,-\frac{1}{2}}$ will couple to the $W$ and $Z$ bosons.  On the other hand, $\chi_{1,0}$ ($=\chi_{1,0}^0$), being uncharged under both $SU(2)$ and $U(1)$, will not couple to the $\gamma$, $Z$, or $W$.  Lastly, $\chi_{3,0}$ contains two charged and one neutral state,
\begin{eqnarray}
\chi_{3,0} &=& \left(\begin{array}{c} \chi_{3,0}^+ \\ \chi_{3,0}^0 \\ \chi_{3,0}^-  \end{array} \right).
\end{eqnarray}
While all three states of $\chi_{3,0}$ will couple to the $W$, we note that $\chi_{3,0}^0$, having both $I_{(3)}$ and hypercharge $0$, will couple to neither the $\gamma$ nor the $Z$.

Next, we explore the Yukawa couplings of these fermions.  First, we write down the couplings which involve $\chi_{1,0}$,
\begin{equation}
{\cal L}\supset - y_{+1} \left(\chi_{2,+\frac{1}{2}} \mbox{ }\epsilon \mbox{ }\tilde{H} \mbox{ }\chi_{1,0}\right) - y_{-1} \left(\chi_{2,-\frac{1}{2}} \mbox{ }\epsilon \mbox{ }H \mbox{ }\chi_{1,0}\right).
\end{equation}
where we have labelled the Yukawa couplings $y_{ij}$ according to the sign of the hypercharge of the doublet $\chi_{2\pm\frac{1}{2}}$ and the subscript $1$ refers to the number of components in the hypercharge-$0$ multiplet.  $\tilde{H}$ denotes $\epsilon H^*$ where $\epsilon= i\tau_2$ and $\tau_2$ is a Pauli matrix.  After electroweak symmetry breaking, $H\rightarrow \frac{1}{\sqrt{2}} \left(\begin{array}{c} 0 \\ v \end{array}\right)$ and $\tilde{H}\rightarrow \frac{1}{\sqrt{2}} \left(\begin{array}{c} v \\ 0 \end{array}\right)$, and we obtain the mass terms
\begin{equation}
\label{eq:singmass}
{\cal L}\supset  -(-1)\frac{y_{+1}v}{\sqrt{2}} \chi_{2,+\frac{1}{2}}^0 \chi_{1,0}-\frac{y_{-1}v}{\sqrt{2}} \chi_{2,-\frac{1}{2}}^0 \chi_{1,0}.
\end{equation}
The Yukawa couplings involving the isospin triplet $\chi_{3,0}$ are slightly more complicated.  We need terms of the form 
\begin{eqnarray}
-y_{+3}& \mbox{ }&\chi_{2,+\frac{1}{2}} \times \tilde{H} \times \chi_{3,0},  \nonumber\\
-y_{-3}& \mbox{ }&\chi_{2,-\frac{1}{2}} \times H \times \chi_{3,0} ,
\end{eqnarray}
which form isospin singlets.  These give mass terms\footnote{The form of these mass terms is simple to see using the following considerations.  Our mass terms contain two isodoublets and one isospin triplet.  Two isospin doublets $\chi=\left(\begin{array}{c} \chi_{\uparrow} \\ \chi_{\downarrow} \end{array} \right)$ and $\chi'=\left(\begin{array}{c} \chi'_{\uparrow} \\ \chi'_{\downarrow} \end{array}\right)$ can be combined into the triplet object $\left(\begin{array}{c} \chi_{\uparrow}\chi'_{\uparrow} \\ \frac{\chi_{\uparrow}\chi'_{\downarrow} + \chi_{\downarrow}\chi'_{\uparrow}}{\sqrt{2}} \\ \chi_{\downarrow}\chi'_{\downarrow} \end{array}\right)$.  The Clebsch-Gordon coefficients \cite{pdg:2012} to combine two $SU(2)$ triplets into a singlet are $\frac{1}{\sqrt{3}}, -\frac{1}{\sqrt{3}}, \frac{1}{\sqrt{3}}$.  Using these relations, normalizing the Yukawa couplings to remove the factor of $\frac{1}{\sqrt{3}}$ and setting $H$ and $\tilde{H}$ to their vacuum expectation values gives Eq. (\ref{eq:tripmass}).}
\begin{eqnarray}
\label{eq:tripmass}
{\cal L}\supset -\frac{y_{+3}v}{\sqrt{2}}\left(\chi_{2,+\frac{1}{2}}^+ \chi_{3,0}^- - \frac{1}{\sqrt{2}}\chi_{2,+\frac{1}{2}}^0 \chi_{3,0}^0\right)-\frac{y_{-3}v}{\sqrt{2}}\left(-\frac{1}{\sqrt{2}}\chi_{2,-\frac{1}{2}}^0 \chi_{3,0}^0 + \chi_{2,-\frac{1}{2}}^-\chi_{3,0}^+\right).
\end{eqnarray}
Thus, our model contains four massive particles.  We have two charged particles, with masses $\frac{|y_{+3}|v}{\sqrt{2}}$ and $\frac{|y_{-3}|v}{\sqrt{2}}$, respectively, and two neutral states, whose masses are determined by the mass matrix $M$ in
\begin{equation}
\label{eq:neutmass}
\left(\chi_{2,+\frac{1}{2}}^0  \chi_{2,-\frac{1}{2}}^0 \right) M  \left( \begin{array}{c} \chi_{1,0}\\ \chi_{3,0}^0 \end{array} \right) = \frac{v}{\sqrt{2}}\left(\chi_{2,+\frac{1}{2}}^0  \chi_{2,-\frac{1}{2}}^0 \right) \left(\begin{array}{cc} -y_{+1} & \frac{-y_{+3}}{\sqrt{2}} \\ y_{-1} & \frac{-y_{-3}}{\sqrt{2}}  \end{array} \right) \left( \begin{array}{c} \chi_{1,0}\\ \chi_{3,0}^0 \end{array} \right).
\end{equation}
{\it A priori}, the Yukawa couplings in Eq. (\ref{eq:neutmass}) are arbitrary complex numbers.  However, we note that we can remove three of the phases in these Yukawa couplings via $U(1)$ rotations on the multiplets.  To see this, we write Eq. (\ref{eq:neutmass}) in terms of the mass eigenstates, which are related to the neutral fields in the polyplet via unitary transformations.  We write
\begin{equation}
\left( \begin{array}{cc} N_{1} & N_{2} \end{array} \right) =    \left( \begin{array}{cc} \chi_{2,\frac{1}{2}}^0 & \chi_{2,-\frac{1}{2}}^0 \end{array} \right) V^{\dagger},
\end{equation}
and 
\begin{equation}
\left( \begin{array}{c} N_{1}' \\ N_{2}' \end{array} \right) =  U  \left( \begin{array}{c} \chi_{1,0}\\ \chi_{3,0}^0 \end{array} \right),
\end{equation}
where $U$ and $V$ are unitary matrices.  We will write $U$ in the form
\begin{equation}
U = e^{i\theta_U} \left( \begin{array}{cc} \alpha_U & \beta_U\\ -\beta^*_U & \alpha^*_U \end{array}  \right),
\end{equation}
where $|\alpha|^2+|\beta|^2=1$.  Via a redefinition of $\theta_U$, we will in turn write this as
\begin{equation}
U = e^{i\theta_U} \left( \begin{array}{cc} 1 & 0\\ 0 & e^{-i\phi_{U1}}  \end{array}\right) \left( \begin{array}{cc} c_U & s_U \\ -s_U & c_U \end{array}  \right)\left( \begin{array}{cc} 1 & 0\\ 0 & e^{-i\phi_{U2}}  \end{array}\right),
\end{equation} 
where $s_U$ and $c_U$ are real and $s_U^2+c_U^2=1$.  Similarly, we will express $V$ as 
\begin{equation}
V = e^{i\theta_V} \left( \begin{array}{cc} 1 & 0\\ 0 & e^{-i\phi_{V1}}  \end{array}\right) \left( \begin{array}{cc} c_V & s_V \\ -s_V & c_V \end{array}  \right)\left( \begin{array}{cc} 1 & 0\\ 0 & e^{-i\phi_{V2}}  \end{array}\right).
\end{equation}  
Then, writing the mass terms as
\begin{equation}
\left( \begin{array}{cc} N_{1} & N_{2}  \end{array} \right)\left(\begin{array}{cc} m_1 & 0\\ 0 & m_2\end{array} \right) \left( \begin{array}{c} N_{1}' \\ N_{2}' \end{array} \right),
\end{equation}
$M$ in Eq. (\ref{eq:neutmass}) becomes
\begin{equation}
\label{eq:M}
M =
e^{i\theta}\left( \begin{array}{cc} 1 & 0\\ 0 & e^{i\phi_{V2}}  \end{array}\right) \left( \begin{array}{cc} c_V & -s_V \\ s_V & c_V \end{array}  \right) \left( \begin{array}{cc} 1 & 0\\ 0 & e^{i\alpha}  \end{array}\right)\left(\begin{array}{cc} m_1 & 0\\ 0 & m_2\end{array} \right)\left( \begin{array}{cc} c_U & s_U \\ -s_U & c_U \end{array}  \right)\left( \begin{array}{cc} 1 & 0\\ 0 & e^{-i\phi_{U2}}  \end{array}\right),
\end{equation}
where we have absorbed $\theta_U$ and $\theta_V$ into a single phase $\theta$ and similarly combined the $\theta_{V1}$ and $\theta_{U1}$ phases into a single angle $\alpha$.  By performing field redefinitions on our multiplets, we can absorb the $e^{i\theta}$, $e^{i\phi_{V2}}$, and $e^{i\phi_{U2}}$ factors into the fields.  (Note that we do the field redefinition on the entire multiplet, and not just on the neutral component, to avoid introducing unnecessary phases in charged-current couplings.)  Thus, the mass matrix can be parameterized by five quantities, $m_1$, $m_2$, $c_V$, $c_U$, and $\alpha$.

For our purposes, however, it is more useful to retain $e^{i\theta}$, $e^{i\phi_{V2}}$, and $e^{i\phi_{U2}}$ and use them to write the mass matrix in a more convenient form.  In order to have standard mass terms for the charged states in Eq. (\ref{eq:tripmass}), we wish to have $y_{+3}$ and $y_{-3}$ real and positive.  Additionally, for convenience, we will define our fields such that $y_{+1}$ is real and negative, and hence the only complex phase in the Yukawa matrix in Eq. (\ref{eq:neutmass}) is contained in $y_{-1}$.  This is accomplished by keeping $M$ of the form in Eq. (\ref{eq:M}), but now $\theta$, $\phi_{V2}$ and $\phi_{U2}$ are functions of the other five parameters; they will fall out of couplings and thus will not contribute to any of the physical quantities discussed below.  

Now that we have a general form for our mass matrix, we wish to verify that we can choose Yukawa couplings such that the lightest particle in our polyplet is neutral in order to have a possibly viable DM candidate.  To see this, let us take the simple case $-y_{+1}=y_{+3}/\sqrt{2}=y_{-3}/\sqrt{2}=1$, $y_{-1}=e^{i\phi}$, where $\phi$ is a phase.  In this case, the two charged states each have mass $v$, while the neutral states have mass squared of $v^2(1\pm\sqrt{1-\frac{1}{2}(1-\cos\phi)})$.  For $\phi=\pi$, this gives degenerate neutral particles with mass $v$.  However, for $\phi=0$, we obtain $m_1=0$, $m_2=\sqrt{2}v$.  By perturbing about this latter solution, we can make the lighter of the neutral states massive yet still lighter than both charged states.  Of course, one can also consider more general Yukawa couplings than those given in this simple example. 

We now consider the couplings of our particles to the SM gauge bosons.  We first note that the $Z$ does not couple to $\chi_{1,0}$ or $\chi_{3,0}^0$ as both have $Y=I_{(3)}=0$.  The $Z$ does, however, couple to the charged states, as well as the neutral components of $\chi_{2,\frac{1}{2}}$ and $\chi_{2,-\frac{1}{2}}$:
\begin{eqnarray}
\label{eq:zcoup}
{\cal L}\supset \frac{g}{\cos\theta_W} Z_{\mu}[&&  \chi_{2,\frac{1}{2}}^{+\dagger} \bar{\sigma}^{\mu} \chi_{2,\frac{1}{2}}^{+}\left(\frac{1}{2}-\sin^2_W\right)   \nonumber +  \chi_{2,\frac{1}{2}}^{0\dagger} \bar{\sigma}^{\mu} \chi_{2,\frac{1}{2}}^{0}\left(-\frac{1}{2}\right) \nonumber\\
+&&  \chi_{2,-\frac{1}{2}}^{0\dagger} \bar{\sigma}^{\mu} \chi_{2,-\frac{1}{2}}^{0}\left(\frac{1}{2}\right) \nonumber+  \chi_{2,-\frac{1}{2}}^{-\dagger} \bar{\sigma}^{\mu} \chi_{2,-\frac{1}{2}}^{-}\left(-\frac{1}{2}+\sin^2_W\right) \nonumber\\
+&&  \chi_{3,0}^{+\dagger} \bar{\sigma}^{\mu} \chi_{3,0}^{+} \left(1-\sin^2\theta_W \right) + \chi_{3,0}^{-\dagger} \bar{\sigma}^{\mu} \chi_{3,0}^{-} \left(-1+\sin^2\theta_W \right)],     
\end{eqnarray}
where $\theta_W$ is the weak mixing angle, $\bar{\sigma}^{\mu}=(1,-\overrightarrow{\sigma})$ and $\overrightarrow{\sigma}$ is the set of Pauli matrices.  We note that the couplings of the neutral fields are equal in magnitude but opposite in sign; we will see below that this allows the coupling of the $Z$ to the neutral mass eigenstates to be somewhat tunable and allows the $Z$ to couple to states which are off-diagonal in the mass basis.

The charged-current interactions of the doublets $\chi_{2,\frac{1}{2}}$ and $\chi_{2,-\frac{1}{2}}$ take a form similar to that of the left-handed fermion doublets in the SM,
\begin{equation}
\label{eq:wcoup1}
{\cal L}\supset \frac{g}{\sqrt{2}} \left[W^+_{\mu}(\chi_{2,\frac{1}{2}}^{+\dagger} \bar{\sigma}^{\mu} \chi_{2,\frac{1}{2}}^{0} + \chi_{2,-\frac{1}{2}}^{0\dagger} \bar{\sigma}^{\mu} \chi_{2,-\frac{1}{2}}^{-})    +W^-_{\mu}(\chi_{2,\frac{1}{2}}^{0\dagger} \bar{\sigma}^{\mu} \chi_{2,\frac{1}{2}}^{+} + \chi_{2,-\frac{1}{2}}^{-\dagger} \bar{\sigma}^{\mu} \chi_{2,-\frac{1}{2}}^{0})\right],  
\end{equation}  
while our $SU(2)$ triplet field interacts via
\begin{equation}
\label{eq:wcoup2}
{\cal L}\supset g\left[W^+_{\mu}(\chi_{3,0}^{0\dagger} \bar{\sigma}^{\mu} \chi_{3,0}^{-} + \chi_{3,0}^{+\dagger} \bar{\sigma}^{\mu} \chi_{3,0}^{0}) + W^-_{\mu}(\chi_{3,0}^{-\dagger} \bar{\sigma}^{\mu} \chi_{3,0}^{0} + \chi_{3,0}^{0\dagger} \bar{\sigma}^{\mu} \chi_{3,0}^{+})   \right].
\end{equation}

We now re-express Eqs. (\ref{eq:zcoup}-\ref{eq:wcoup2}) in terms of the mass eigenstates.  For the neutral states, we combine $N_1$ and $N_1'$ into a single massive Dirac spinor with $N_1$ as its left-handed component.  This Dirac spinor has mass $m_1$ and we will now refer to it as $N_1$.  Analogous notation applies to $N_2$.  Without loss of generality, we will assume that $m_2\ge m_1$.  The couplings of these particles to the $Z$ boson are
\begin{eqnarray}
\label{eq:zcoupm}
{\cal L}\supset \frac{g}{\cos\theta_W} Z_{\mu}\left(\frac{1}{2}\right)[&&(s_V^2-c_V^2)(\bar{N}_1 \gamma^{\mu}P_L N_1-\bar{N}_2 \gamma^{\mu}P_L N_2)\nonumber\\ + &&2s_Vc_V(e^{-i\alpha}\bar{N}_1 \gamma^{\mu}P_L N_2 + e^{i\alpha}\bar{N}_2 \gamma^{\mu}P_L N_1)],
\end{eqnarray}
where $P_L$ is the left-handed projection operator.  Here we see that the mass-diagonal couplings of $N_1$ and $N_2$ to the $Z$ can be tuned to be small if $|c_V|\approx |s_V|$ and that the off-mass-diagonal coupling can be substantial.

Similarly, the two massive charged particles also interact with the $Z$.  Denoting the particle with mass $\frac{y_{+3}v}{\sqrt{2}}$ as $\chi_1$ and the particle with mass $\frac{y_{-3}v}{\sqrt{2}}$ as $\chi_2$, their couplings to the $Z$ are

\begin{eqnarray}
{\cal L}\supset \frac{g}{\cos\theta_W} Z_{\mu} [&& \bar{\chi_1} \gamma^{\mu} \left[ \left(\frac{1}{2}-\sin^2\theta_W \right)P_L + \left(1-\sin^2\theta_W \right)P_R \right]\chi_1\nonumber\\
+&& \bar{\chi_2} \gamma^{\mu} \left[ \left(-\frac{1}{2}+\sin^2\theta_W \right)P_L + \left(-1+\sin^2\theta_W \right)P_R \right]\chi_2].
\end{eqnarray}
Note that, with our definitions, $\chi_1$ is positively charged and $\chi_2$ is negatively charged.

In the mass basis, the charged-current couplings in Eqs. (\ref{eq:wcoup1}) and (\ref{eq:wcoup2}) become
\begin{equation}
\label{eq:wcoup1m}
{\cal L}\supset \frac{g}{\sqrt{2}}W^+_{\mu}\left[\bar{\chi}_1 \gamma^{\mu}P_L (c_V N_1-s_V e^{-i\alpha}N_2) +(s_V\bar{N}_1+c_Ve^{i\alpha}\bar{N}_2)\gamma^{\mu}P_L\chi_2   \right]+\mbox{h.c.},
\end{equation}
and
\begin{equation}
\label{eq:wcoup1m2}
{\cal L}\supset g W^+_{\mu}\left[- \bar{\chi}_1\gamma^{\mu}P_R(s_U N_1 +c_U N_2)   -(s_U\bar{N}_1+c_U\bar{N}_2)\gamma^{\mu}P_R \chi_2  \right]+\mbox{h.c.},
\end{equation}
respectively.

Now that we have established the couplings of the particles in our polyplet, we can discuss some of the constraints on this model.

\section{Constraints}
\label{sec:const}
As this model contains new weak-scale fermions which couple to the $W^{\pm}$ and $Z$ bosons, it is important to examine existing constraints from colliders.  A basic constraint comes from the measurement of the invisible width of the $Z$ at LEP \cite{Abreu:1991pr}, which requires that a new stable neutrino with SM couplings to the Z have a mass of at least $45.0$ GeV.  We thus expect that our lightest neutral particle would have a mass of at least this magnitude, although this constraint could be evaded if only one neutral particle was light and $|s_V^2-c_V^2|$ was sufficiently small.

There also exist a few constraints from colliders on new neutral particles with masses greater than $M_Z/2$. The authors of \cite{Fox:2011fx} use the single-photon spectrum from DELPHI \cite{Abdallah:2003np,:2008zg} to constrain the production of invisible particles in $e^+e^-$ collisions via either effective operators or on-shell mediators.  For the case most relevant here, production of $\chi_1$ of mass $>M_Z/2$ via a $Z$ boson, however, the limits are weaker than those from direct detection which will be discussed in Sec. \ref{sec:dm}.  Similar conclusions apply to studies of mono-jet events at the Tevatron \cite{Bai:2010hh}. 

It is also possible that the heavier of our two neutral states, $N_2$, could be produced at colliders via $Z\rightarrow N_2 N_1$ or $Z\rightarrow N_2 N_2$.  If the off-diagonal $Z$ couplings are not tiny, $N_2$ could decay down to $N_1$ via $N_2\rightarrow N_1 Z$.  (Also, if $N_2$ is heavier than $\chi_1$ or $\chi_2$, it could decay to $\chi_{1,2}W^{\pm}$, followed by the subsequent decay of the $\chi_{1,2}$.)  The similar decay of fourth-generation neutrinos with both Dirac and Majorana masses has been investigated in \cite{Carpenter:2010sm}, which used LEP SUSY searches \cite{Achard:2003ge} to conclude that LEP would likely exclude masses of the heavier neutrino below $\approx 100$ GeV, and possibly as high as $130$ GeV, if the lighter of the two states had a mass of $\sim 50$ GeV.  Although the couplings in our case are somewhat different, we would expect the limits here to be roughly similar.  For a related analysis at CDF which gives weak constraints, see \cite{CDF:2011ah}. 

There also exist collider limits on heavy charged leptons.  L3 \cite{Achard:2001qw} considered the case of a new charged lepton $L^{\pm}$ which decays to a heavy neutral lepton $L^0$ and a $W^{\pm}$, assuming that the $L^0$ weighs at least $40$ GeV and that the mass difference between the $L^{\pm}$ and the $L^0$ was between $5$ and $60$ GeV; they place lower bounds on the $L^{\pm}$ mass of $\sim 100$ GeV.  As in the case of heavy neutral leptons, the constraints that these limits place on our model are dependent on the mixing angles, but we would nonetheless expect these constraints to roughly hold.

Because models with extra fermion generations often have signals which are boosted relative to the SM, one may worry that this model would be ruled out by the recent discovery of a $\sim 125$-GeV Higgslike particle at LHC \cite{:2012gk,:2012gu} or by previous Higgs searches at LHC \cite{Chatrchyan:2012tx,ATLAS:2012ae} and at the Tevatron \cite{Benjamin:2011sv}.  (For a recent survey of the experimental limits on such scenarios, see \cite{Gunion:2011ww}.)  However, constraints on fourth-generation scenarios are not directly applicable here;  this model contains no heavy colored particles beyond the SM top quark; therefore, there is no enhancement in Higgs production via gluon fusion in this model.  This model does give a new contribution to the decay $H\rightarrow \gamma\gamma$ via fermion loops which is opposite in sign to the main ($W$-loop) SM contribution, thus reducing the branching fraction of $H\rightarrow \gamma\gamma$ relative to the SM.  As the early results from LHC indicate a possible enhancement in $2$-photon decays \cite{:2012gk,:2012gu}, this may imply a tension with our model if the Higgs sector is eventually confirmed to consist of a single $\sim 125$-GeV resonance.  This issue does not arise, however, if the $125$-GeV resonance turns out to not be a Higgs boson.  Additionally, this tension may be relieved if the $125$-GeV resonance is just the lightest of multiple Higgs scalars; if a heavy Higgs can decay to non-SM particles, then the current exclusion bounds for large Higgs masses can be loosened, similar to the case of Higgs decays to fourth-generation neutrinos in \cite{Carpenter:2011wb}. We also point out that in more general polyplet scenarios with larger $n$, $H\rightarrow \gamma\gamma$ could be enhanced.

In addition to constraints from direct searches, however, we may consider the implications of heavy fermions on electroweak precision parameters.   Thus, here we consider the contributions of our new fermions to the oblique parameters $S$ and $T$ \cite{Peskin:1991sw}.  While a full exploration of the parameter space is beyond the scope of this work, we consider a special case, with a mass matrix of the form 
\begin{equation}
M= \frac{v}{\sqrt{2}}\left(\begin{array}{cc} -y_{+1} & \frac{-y_{+3}}{\sqrt{2}} \\ y_{-1} & \frac{-y_{-3}}{\sqrt{2}}  \end{array} \right) = \frac{1}{\sqrt{2}}\left(\begin{array}{cc} -m_1 & -m_2\\ m_1 & -m_2\end{array} \right),
\end{equation} 
which gives $N_1$ a mass of $m_1$ and $N_2$, $\chi_1$, and $\chi_2$ a mass of $m_2$.  Diagonalizing this mass matrix leads to the mixing parameters 
\begin{eqnarray}
\label{eq:simpcoup}
s_V&=&\frac{1}{\sqrt{2}},\nonumber\\
c_V&=&\frac{-1}{\sqrt{2}},\\
c_U&=&1,\nonumber\\
\alpha&=&0.\nonumber
\end{eqnarray}
Inspecting Eqs.(\ref{eq:zcoupm}), we see that this scenario yields no mass-diagonal $Z$ couplings to $N_1$ and $N_2$.  This feature will be particularly useful when we consider the potential of $N_1$ to be a dark matter candidate in Sec. \ref{sec:dm}.  Additionally, from Eqs. (\ref{eq:wcoup1m}) and (\ref{eq:wcoup1m2}), we see that the charged-current couplings of $N_1$ are entirely left-handed.  

\begin{figure}
\epsfig{file=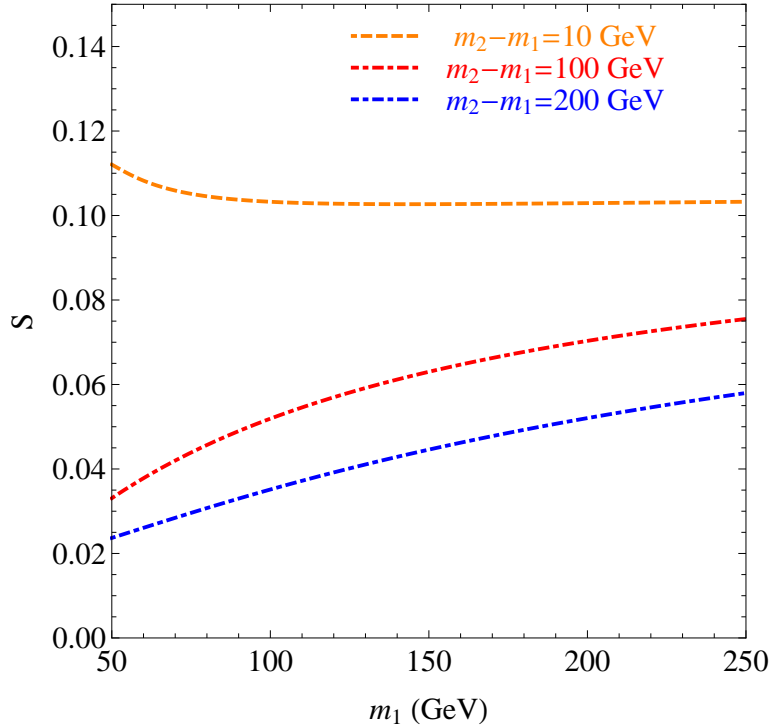,width=4in}
\caption{The contribution to $S$ in the simplified mass and mixing scenario as a function of $m_1$ for three values of the mass difference $m_2-m_1$.}
\label{fig:sparam}
\end{figure}

In Fig.(\ref{fig:sparam}), we plot $S$ as a function of $m_1$ for this simplified scenario of masses and mixings.  Here, we take $m_2=m_1+10, 100, 200\mbox{ GeV}$.  We see that $S$ is typically ${\cal O}(0.01-0.1)$ in all three cases.  We note that in the case that $m_1=m_2>>M_Z$, $S=1/(3\pi)$.  We compare this with the results of \cite{Bai:2011aa}\footnote{Also see \cite{Kribs:2007nz} for a discussion of the oblique parameters in the context of fourth-generation scenarios.} which uses results from the Gfitter Group \cite{Baak:2011ze} to find that the new physics contributions to $S$ and $T$ are constrained to be
\begin{eqnarray}
\label{eq:stnum}
S|_{U=0}&=&0.07\pm0.09-\frac{1}{12\pi}\ln{\frac{m_h^2}{(120\mbox{GeV})^2}},\nonumber\\
T|_{U=0}&=&0.10\pm0.08+\frac{3}{16\pi\cos^2\theta_W}\ln{\frac{m_h^2}{(120\mbox{GeV})^2}}.
\end{eqnarray}
Additionally, for general $m_1$, $m_2$ in this simplified scenario, $T=0$; this is because the Yukawa couplings do not break the custodial symmetry for the case $y_{+1}=y_{-1}$, $y_{+3}=y_{-3}$.\footnote{We can see this by rewriting our fields as $\chi_{3,0}=\left(\begin{array}{cc}\frac{\chi_{3,0}^0}{\sqrt{2}} & -\chi_{3,0}^+\\\chi_{3,0}^- & -\frac{\chi_{3,0}^0}{\sqrt{2}}\end{array}\right)$, $\chi_{2}=\left(\begin{array}{cc}\chi_{-\frac{1}{2}}^0 & -\chi_{\frac{1}{2}}^+\\\chi_{-\frac{1}{2}}^- & -\chi_{\frac{1}{2}}^0\end{array}\right)$, $H=\left(\begin{array}{cc} h^{0*}&h^+ \\-h^- & h^0\end{array}\right)$.  Under $SU(2)_L$ and $SU(2)_R$ rotations $U_L$ and $U_R$, $\chi_{3,0}\rightarrow U_L\chi_{3,0}U_L^{\dagger}$, $\chi_{2}\rightarrow U_L\chi_{2}U_R^{\dagger}$, $H\rightarrow U_L H U_R^{\dagger}$.  Yukawa couplings in this scenario can be written as $\mbox{Tr}[\chi_{3,0}\chi_{2}H^{\dagger}]$ and $\mbox{Tr}[\chi_{2}H^{\dagger}]\chi_{1,0}$; when the $H$ vacuum expectation value breaks $SU(2)_L\times SU(2)_R$ down to the custodial symmetry, $U_L=U_R$, these terms remain invariant.}  (Outside of this specific scenario, however, $T$ is not generally zero.)  We thus see that this model can easily produce values for $S$ and $T$ which are in reasonable agreement with experiment for $m_H\sim 125$ GeV.  We will return to this in Sec. \ref{sec:dm} when we discuss the interplay between these results and constraints from direct detection.

Lastly, we briefly mention a few other possible constraints on the polyplet model.  First, as the couplings of our new fields contain a complex phase $\alpha$, it is possible that our model contains contributions to CP-violating observables, such as electric dipole moments (EDMs) of SM particles (constraints on which can be found in \cite{pdg:2012}), once we move beyond tree-level calculations.  We note, however, that $\alpha$ can be chosen to be small; we thus do not consider constraints from EDMs here, but they could be of use in a more specific model.  Also, as has been pointed out in \cite{Cheung:2012nb}, the addition of new heavy fermions to the SM can have ramifications for vacuum stability; here, we simply assume that this issue will be resolved by some other new physics added to the SM.

We now move on to discuss the viability of $N_1$ as a DM candidate.

\section{$N_1$ as dark matter}
\label{sec:dm}
As the lightest member of our polyplet, $N_1$ is stable and therefore a potential dark matter candidate.  (For large $n$, this would be a result of there being no dimension-four $SU(2)$-invariant operators which couple the members of our polyplet to the SM fermions; for $n=2$, it is enforced by our choice of pseudo-lepton number.)  In order for $N_1$ to possibly be a realistic DM candidate, it must fulfill two basic requirements.  First, it must annihilate to SM particles in the early universe sufficiently to have a relic density which is equal to or less than the observed DM density today, and, second, it must interact with nucleons sufficiently weakly that it evades current constraints from direct detection.  Both of these processes are mediated via interactions of $N_1$ with the $Z$ boson, the Higgs boson, or a combination of the two.

In trying to simultaneously fulfill the relic density and direct detection requirements, a potential conflict arises.  If DM is a thermal relic, its observed density of $\Omega_{DM}h^2=0.112\pm 0.006$ implies a thermally-averaged annihilation cross-section of $<\sigma |v|>\sim 3\times 10^{-26}\mbox{ cm}^3/\mbox{s}$.  (For a general review of the status of dark matter and its detection see, for example, \cite{pdg:2012}.)  The observation that this cross-section is typical of weak-scale processes is commonly referred to as the WIMP miracle.  However, direct detection experiments have put constraints on DM-nucleon cross-sections as stringent as  ${\cal O}(2\times 10^{-45}\mbox{cm}^2)\sim {\cal O}(5\times 10^{-18}\mbox{ GeV}^{-2})$ \cite{Aprile:2012nq,Ahmed:2011gh}.  If DM-nucleon interactions are mediated by physics at a new scale $\Lambda$, DM-nucleon cross-sections would typically be of order $\mu^2/\Lambda^4$ \cite{Jungman:1995df}, where $\mu$ is the DM-nucleon reduced mass, ${\cal O}(1\mbox{ GeV})$.  From this, one can see that a value of $\Lambda\sim 100$ GeV or even $\Lambda \sim 1$ TeV would produce a DM-nucleon cross-section which is ruled out by direct-detection experiments by several orders of magnitude.  Thus, in order to have a DM candidate which simultaneously has an acceptable relic density and evades the direct detection bounds, there must be some significant difference between the physics which controls the DM annihilation and that which dominates DM scattering with nucleons. 

As a first example, we consider the case of a relatively light $N_1$, $m_1<m_W$,where the $N_1$ can only annihilate to the light SM fermions.  As long as $s_V^2-c_V^2$ is not very small (and annihilation via a $H$ is non-resonant), annihilation via a $Z$ will dominate over $H$-mediated processes.   If we take the $Z$ coupling to be full-strength,  $|s_V^2-c_V^2|\sim 1$, $N_1$ will comprise a subdominant component of DM, yet will fail to evade direct detection constraints.  Also, as long as both DM annihilation and nucleon scattering are dominated by diagrams mediated by a $Z$ boson, adjusting the coupling of $N_1$ to the $Z$ will not help mitigate this conflict.  Although reducing  $s_V^2-c_V^2$ will reduce the effective coupling between the DM and nucleons, it will similarly reduce the annihilation cross-section, thus increasing the current-day relic density; this effect will compensate for the reduced coupling, leaving the direct-detection constraints largely unchanged.  Therefore, we must have non-negligible contributions to $N_1$ annihilation from processes mediated by the $H$.

If we keep $m_1<M_W$ but decrease the mass-diagonal coupling between the $Z$ and $N_1$ to zero such as in the simplified scenario we explored in the discussion on oblique parameters in Sec. \ref{sec:const}, we can have $N_1$ interact with the SM entirely via the Higgs.  In this case, one might hope that the direct-detection cross-section will be sufficiently suppressed by light quark Yukawa couplings to evade the relevant constraints.  If $m_1$ is small, however, $N_1$ can only annihilate to the light SM fermions, and, thus, the dominant piece of this cross-section is suppressed by the Yukawa coupling of the $b$ quark.   Unless $m_1 \sim m_H/2$, this annihilation cross-section is too small to yield a reasonable relic density.  Additionally, if one chooses $m_1$ to obtain the correct relic density, direct detection bounds still rule out this scenario for a light $N_1$.  The direct detection cross-section is dominated by the $s$-quark contribution to the nucleon,
\begin{equation}
\label{eq:hsnuc}
\sigma_{nucleon-N_1}\approx\frac{\mu^2}{\pi}B^2_{Ns}\left(\frac{m_1 m_s}{v^2 m_H^2}\right)^2,
\end{equation}
where $B_{Ns}=3.36\pm1.45$ is the factor which converts between quark-level and nucleon-level effective operators; we take this value from \cite{Bai:2010hh} which used $m_s=105\pm 25$ MeV.  (We note that the error bar on this factor is large and different conversion factors are used in the literature; see, for example, \cite{Agrawal:2010fh,Belanger:2008sj}.  Additionally, the expected contributions from $u$, $d$, and heavy quarks are smaller than or comparable to the error on the value in Eq. (\ref{eq:hsnuc}).)  This yields
%\begin{equation}
%\sigma_{nucleon-N_1}\approx (3.7\pm 2.7)\times 10^{-43}\mbox{ cm}^2 \left(\frac{m_1}{100\mbox{ GeV}}\right)^2 \left(\frac{100\mbox{ GeV}}{m_H}\right)^4
%\label{eq:hsnucnum}
%\end{equation}
\begin{equation}
\sigma_{nucleon-N_1}\approx 3.7\times 10^{-43}\mbox{ cm}^2 \left(\frac{m_1}{100\mbox{ GeV}}\right)^2 \left(\frac{100\mbox{ GeV}}{m_H}\right)^4
\label{eq:hsnucnum}
\end{equation}
with an error bar of the same order.  As the limits on the spin-independent DM cross-section from Xenon100 \cite{Aprile:2012nq} are ${\cal O}(2\times 10^{-45}\mbox{ cm}^2)$, this scenario is ruled out by approximately an order of magnitude for low Higgs masses. 

\begin{figure}[h]
\epsfig{file=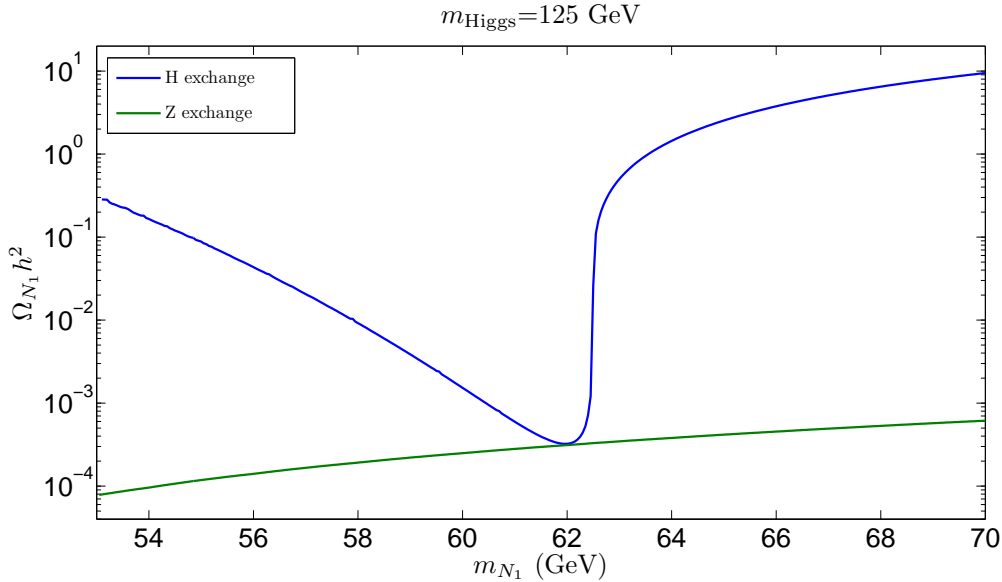,width=6in}
\caption{$\Omega_{N_1} h^2$ as a function of $m_1$ in the case of resonant annihilation through $H$ for $m_H=125$ GeV.  At best, resonant annihilation can only suppress the relic density to a level comparable to that of annihilation via the $Z$ (also shown).}
\label{fig:reso}
\end{figure}

If we are willing to allow $N_1$ to comprise only a very subdominant component of DM, we can try to use resonant annihilation via the Higgs to make the current relic density of $N_1$ small.  In this case, we tune the $N_1$ mass to be $m_1\sim M_H/2$, but we do not fine-tune the $Z$ coupling to be small.  Although the $N_1$-nucleon cross-section will not be suppressed, one can hope that the reduced $N_1$ relic density will allow $N_1$ to evade the direct detection constraints at the expense of identifying the primary component(s) of DM.   In Fig. \ref{fig:reso}, we show the relic density $\Omega_{N_1} h^2$ of $N_1$ as a function of $m_1$ assuming resonant annihilation through $H$ for $m_H=125$ GeV; we include the effects of thermal averaging similarly to what was done in \cite{Griest:1990kh}.  (See also \cite{Ibe:2008ye,Ibe:2009dx}.)  We see that it is possible to achieve relic densities as low as $\mbox{few}\times 10^{-4}$.  We note, however, that only for points very close to resonance does the resonant annihilation cross-section approach the cross-section for annihilation through a $Z$; the relic density in the case of annihilation via the $Z$ is shown separately in the plot.  Thus, this suppression is inadequate to evade direct detection constraints; the $N_1$-nucleon cross-section via $Z$ exchange calculated for Xenon100 is $\sim 6\times 10^{-40}\mbox{ cm}^2$ , more than $\sim 2$ orders of magnitude too large to be compatible with the results of Xenon100 \cite{Aprile:2012nq}, even given the above relic density suppression. 
%\begin{figure}[h]
%\epsfig{file=sigmaNP_final2.eps,width=6in}
%\caption{The $N_1$-nucleon cross-section weighted by $\Omega_{N_1}/\Omega_{DM}$, calculated for Xenon100, in the case of resonant annihilation to light fermions via $H$. {\color{red}$m_H=125$ GeV.}}
%\label{fig:ressig}
%\end{figure}

\begin{figure}[h]
\epsfig{file=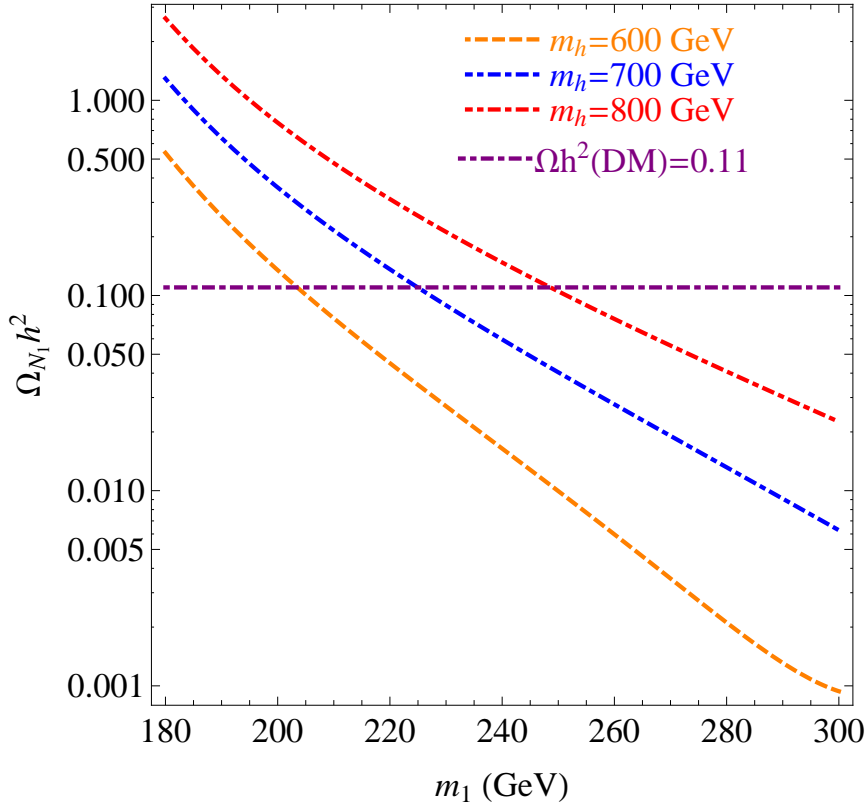,width=4.5in}
\caption{$N_1$ relic density as a function of $m_1$ for $m_H= 600, 700$ and $800$ GeV assuming SM Higgs couplings and taking the other polyplet fields to be heavy.  The horizontal line is $\Omega_{DM}h^2\sim 0.11$.}
\label{fig:higgso}
\end{figure}

\begin{figure}[h]
\epsfig{file=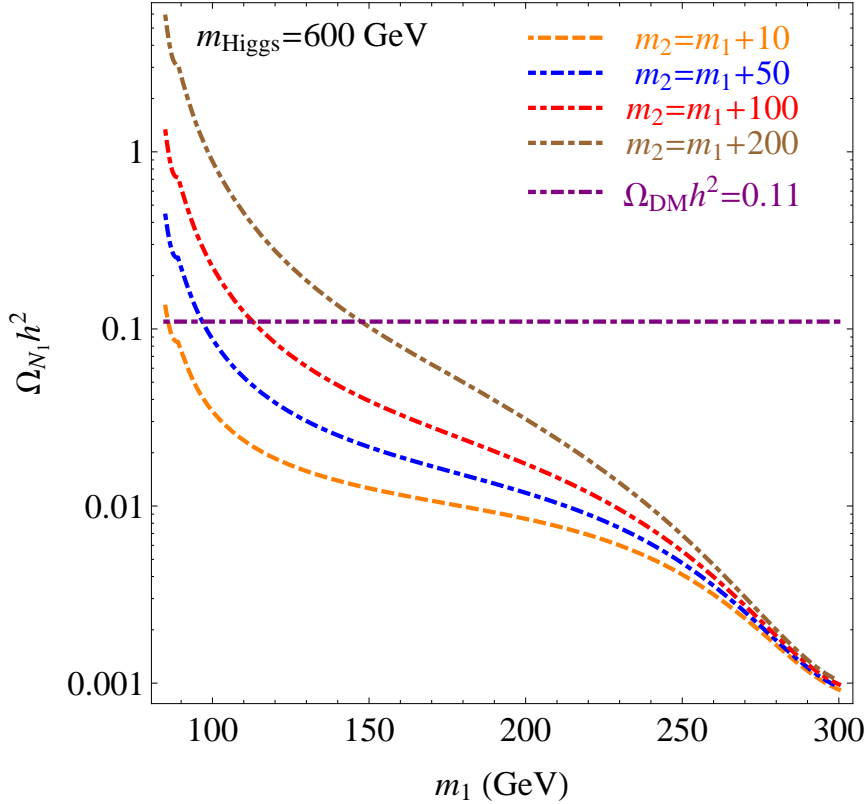,width=4.5in}
\caption{$N_1$ relic density as a function of $m_1$ for various values of $m_2$ and $m_H= 600$ GeV, assuming SM Higgs couplings.  The horizontal line is $\Omega_{DM}h^2\sim 0.11$.}
\label{fig:higgso2}
\end{figure}

A solution is possible, however, for larger $m_1$ if the Higgs sector contains a heavy scalar; this may happen if the Higgs sector contains multiple scalars or if the $125$-GeV particle discovered at LHC turns out to not be closely related to electroweak symmetry breaking.  If we tune the mass-diagonal coupling of $N_1$ to the $Z$ to zero ($s_V^2-c_V^2$ must be $\lesssim \mbox{few}\times 10^{-3}$ to evade direct detection limits), and increase $m_1$ so that annihilations to gauge bosons $N_1 \bar{N}_1\rightarrow H \rightarrow W^+W^-, ZZ$ and $t\bar{t}$ become possible, we can simultaneously fulfill the relic density requirement and evade the constraints from direct detection.  In this case, the annihilation cross-section for $N_1$ is not suppressed by any small Yukawa couplings, while the main contribution to the direct-detection cross-section is still suppressed by the $s$ quark Yukawa, assuming SM Higgs couplings.  This solution does, however, require $N_1$ to couple to a rather heavy Higgs boson, $m_H\sim{\cal O}(600\mbox{ GeV})$.    We plot the relic density of $N_1$ in this case in Fig. \ref{fig:higgso}, for $m_H= 600, 700$ and $800$ GeV, along with the observed DM relic density $\Omega_{DM}h^2\sim 0.11$.  Here, we assume a single Higgs boson with a vacuum expectation value of $246$ GeV; results will differ for multi-Higgs scenarios.  Additionally, we have taken the other polyplet fields to have masses much larger than $m_1$.   For the case where the other polyplet fields have masses comparable to $m_1$, one must also include contributions to $N_1 \bar{N}_1\rightarrow  W^+W^-, ZZ$ via t-channel exchange of $N_2$, $\chi_1$, and $\chi_2$. We plot the relic density of $N_1$ in this case in Fig. \ref{fig:higgso2} for $m_h=600$ GeV and various values of $m_2$, taken to be the common mass of $N_2$, $\chi_1$, and $\chi_2$; for large $m_1$, the s-channel Higgs exchange dominates, and the lines converge to the value in Fig. \ref{fig:higgso}.

For these values of $m_1$ and $m_H$, it can be seen from Eq. (\ref{eq:hsnucnum}) that typical values of the direct detection cross-section are ${\cal O}(10^{-45}\mbox{ cm}^2)$.  Thus, we see that if $m_1\sim {\cal O}(200 \mbox{ GeV})$, $s_V^2-c_V^2\lesssim  \mbox{few}\times 10^{-3}$ and we have a heavy Higgs, $N_1$ can be a successful thermal DM candidate; additionally, the typical $N_1$-nucleon cross-sections are accessible by near-future direct detection experiments.

\begin{figure}[h]
\epsfig{file=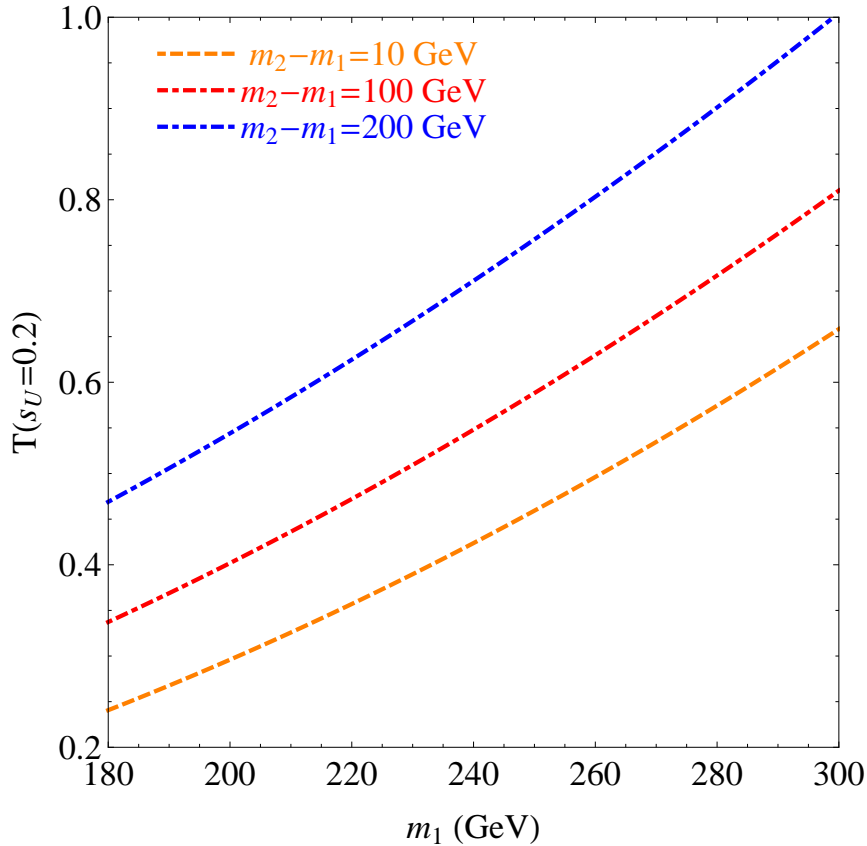,width=4.5in}
\caption{$T$ as a function of $m_1$ for $s_U=0.2$ and three values of $m_2-m_1$.}
\label{fig:tbig}
\end{figure}

We briefly comment on the feasibility of a heavy Higgs boson.  Values of $m_H$ up to $600$ GeV have been ruled out by CMS \cite{Chatrchyan:2012tx}, although it is possible that this constraint could be loosened slightly in the polyplet scenario as new decay modes of the Higgs, $H\rightarrow N_1\bar{N}_1, N_2\bar{N}_2, \chi_1\bar{\chi}_1, \chi_2\bar{\chi}_2$ are available.  A more concerning constraint comes from the oblique parameters.  If we again consider Eq. (\ref{eq:stnum}), we see that (in single-Higgs scenarios) a $600$-$800$ GeV Higgs boson requires that new physics make a contribution to $T$ of approximately $0.35$-$0.4$.  To see that such a contribution can arise in the polyplet scenario, we revisit our simplified couplings in Eq. (\ref{eq:simpcoup}), but we now take $s_U$ to be small but nonzero.    In this case, the masses of the neutral particles are still $m_1$ and $m_2$, but the charged particles now have masses $m_2c_U\pm m_1s_U$.  A plot of $T$ for $s_U=0.2$ is shown in Fig. \ref{fig:tbig}; we see that values $T\sim 0.4$ can be achieved for reasonably small $s_U$.  Of course, even in this heavy-Higgs scenario, a full treatment should also consider contributions to the oblique parameters from the $125$-GeV resonance discovered at LHC.  For this we will have to wait for improved knowledge of its identity and couplings; however, we take our results as indication that the polyplet scenario can accomodate a range of values for the oblique parameters.

As our solution to the dark matter relic density and evasion of the direct detection constraints relied on tuning the $N_1\bar{N}_1Z$ coupling to be close to $0$, one may worry that 1-loop corrections might ruin this solution.  Here, we assume that the fine-tuning of the $N_1\bar{N}_1Z$ coupling includes such corrections.  However, other 1-loop effects could also be relevant.  It is possible that $N_1$ could interact with the SM via a magnetic moment.  We have checked (using the results of \cite{Dvornikov:2004sj}), for the simplified couplings in Eq.  (\ref{eq:simpcoup}), that the magnetic moment of $N_1$ vanishes at one-loop order, due to a cancellation between diagrams containing a $\chi_1$ and those containing a $\chi_2$.  However, we note that the sizes of these individual contributions are of the order of those which would be ruled out by Xenon100 \cite{Aprile:2012nq,Fortin:2011hv}, and, therefore, the ability to evade this bound will be somewhat dependent upon the particular mass and mixing parameters chosen for the model.  It is also possible that $N_1$ could interact with quarks via box diagrams; we assume that these contributions to the direct-detection cross-section are suppressed by ${\cal O}((\alpha/4\pi)^2)$ relative to a weak-scale cross-section. 

Given that our set of parameters for $N_1$ to be a DM candidate involved a heavy Higgs boson, we briefly comment on the scenario in which the Higgs sector is confirmed to consist of a single, $125$-GeV scalar with SM couplings. In this case, constraints from direct detection may require us to consider somewhat more complicated solutions.  We outline a few of these here.

If we are willing to accept additional fine-tuning, we can return to the scenario of resonant annihilation; if the mass-diagonal coupling of $N_1$ to the $Z$ were reduced from its maximal value by approximately  a factor of $30$ or more, constraints from direct detection experiments could be evaded for $m_1\approx m_H/2$.  For $m_1$ very close to $m_H/2$, the $N_1$ relic density would be ${\cal O}(10^{-3})$ the observed DM relic density.

Another possibility is that $N_1$ forms a significant, but subdominant, component of DM.  If we return to the scenario of $m_1<M_W$, with the mass-diagonal coupling between $N_1$ and $Z$ tuned to zero, we see from Eq. (\ref{eq:hsnucnum}) that Xenon100 rules out this scenario if DM is comprised solely of $N_1$.  It does, however, allow for $N_1$ to comprise a significant (of order a few percent) fraction of DM, which can be attained if $N_1$ annihilation via the Higgs has a mild resonant enhancement. %(We also point out, somewhat optimistically, that if the conversion factor between between the $s$ quark- and nucleon-level operators is somewhat smaller than the smallest values given in \cite{Agrawal:2010fh,Belanger:2008sj}, then the $N_1$ relic density could be of order of the observed DM relic density and yet evade direct detection constraints for $m_H \sim 125$ GeV.)

Alternatively, we could accomplish much the same situation using coannihilations; if the mass-diagonal Z coupling were made small, but the $N_1$ was close in mass to one of the other polyplet fields (with a mass difference of order $\sim 10$ GeV or less), we could have the relic density controlled by annihilations via the $Z$ and/or the $W$ without the $Z$ contributing significantly to direct detection.\footnote{One could alternatively introduce a small Majorana mass term for $N_1$, rendering it an inelastic DM \cite{TuckerSmith:2001hy} candidate; this would require breaking our pseudo-lepton-number symmetry, but we could still prevent $N_1$ from decaying to SM particles by requiring lepton and baryon number conservation.  However, we do not pursue this route, as the introduction of an arbitrary mass term not tied to the electroweak scale violates the spirit of the polyplet model.}  As above, $N_1$ would have to be a subdominant (up to several percent) component of DM, in order to evade direct detection constraints from its interaction via a $125$-GeV Higgs (see Eq. (\ref{eq:hsnucnum})), but that this  number could perhaps be somewhat ameliorated to ${\cal O}(10\%)$ if the conversion factor in Eq. (\ref{eq:hsnuc}) takes a particularly small value.  We note that the constraints in \cite{Carpenter:2010sm,CDF:2011ah,Achard:2001qw} disappear in the case that the mass splitting between $N_1$ and the next lightest state is small.

Lastly, another possibility is to allow our polyplet fields to decay to SM particles through the introduction of higher-dimensional operators which break pseudo-lepton number; in this case, we can also relax the requirement that the lightest new particle be neutral.  Although imposing the pseudo-lepton-number symmetry and then breaking it is somewhat {\it ad hoc}, such a situation is much more reasonable for more general polyplet scenarios; if $n$ is large, the DM stability is ensured by an accidental symmetry, instead of the imposition of pseudo-lepton number; in these cases, DM decay would be permitted, but could only occur through higher-dimensional operators.  Of course, in this case, we evade direct detection constraints, but we fail to have a DM candidate.

\section{Conclusions and Outlook}
\label{sec:conc}
In this work, we have explored a simple ``polyplet'' extension of the SM which contains new fermion $SU(2)$ multiplets and which has the potential to generate a viable DM candidate.  This model contains four new Dirac fermions with weak-scale masses and causes no difficulties with triangle anomalies.  We also find that this simple polyplet model is compatible with current collider and electroweak constraints and that a DM candidate is possible if we are willing to accept the possibility of a heavy ($m_H\gtrsim 600$ GeV) Higgs boson.  Additionally, we expect that solutions where $N_1$ is a subdominant but non-negligible (possibly ${\cal O}(10\%)$), component of DM can be found in the case that the 125-GeV particle discovered at LHC is confirmed to be the SM Higgs boson, although additional fine-tuning may be needed.

We briefly comment here on a possible resemblance between the polyplet model and supersymmetric models.  It may be noted that our $n=2$ model contains an $SU(2)$ triplet, two doublets, and a singlet; the same is true of the gaugino and Higgsino sectors of the Minimal Supersymmetric Standard Model (MSSM).  However, several differences exist, the most obvious being the lack of sfermion analogues in the polyplet model.  Also, all masses for the polyplet fermions are Dirac, while those in the MSSM are Majorana. (However, for an example of an R-symmetric supersymmetric model--the MRSSM--with Dirac gauginos, see \cite{Kribs:2007ac}.)  Lastly, here the Yukawa couplings are arbitrary parameters; they are not related to gauge couplings as in supersymmetric scenarios.

This work has many possible extensions.  Although here we have only considered the simplest viable ($n=2$) polyplet model, generalizing to higher $n$ would automatically give the stabilization of DM, which here was done using a pseudo-lepton-number.  Several issues would have to be investigated for a specific model with $n>2$, such as confirming that Yukawa couplings can be chosen such that a neutral member of the polyplet is the lightest and calculating the effects of larger multiplets on the oblique parameters.  Also, while in this work we have insisted that all fields in our polyplet receive their masses from Yukawa couplings with the SM Higgs, the phenomenology would likely change substantially if we were to allow the introduction of Majorana or dimension-three Dirac mass terms.  Additionally, it would be interesting to investigate the signatures of polyplets at LHC.  We leave these explorations for future work.

Although here we have focused on the relevance of a polyplet model to DM, polyplets are interesting in their own right; they are a simple extension of the SM with weak-scale masses and reasonably well-defined couplings, which offers the possibility of a rich phenomenology accessible at current-day experiments.  In polyplet models which do not contain a DM candidate, it may be necessary to introduce (possibly higher-dimensional) operators to allow the polyplet fields to decay down to SM particles, but these operators may be sufficiently suppressed that they have no visible effects at colliders, and the lightest new particle may be long-lived on collider time-scales.  Additionally, it may be interesting to see if the polyplet framework could be used to address other issues, such as neutrino mass.  In the $n=2$ model which we explore here, the $SU(2)$ triplet and both $SU(2)$ doublets can couple to the SM leptons at dimension three or four in the absence of pseudo-lepton-number conservation; all of these can contribute to neutrino mass.  Also, it may be possible to incorporate a polyplet containing an $SU(2)$ triplet into a Type-III seesaw model.

In conclusion, fermions placed into larger multiplets of the SM gauge group can be phenomenologically interesting, yet still undiscovered.  The polyplet framework offers a simple extension of the SM with implications for dark matter and collider experiments and shows promise for future studies.

\section{Acknowledgements}
The authors would like to thank S. Bauman, M. Gonderinger, W.-Y. Keung, M. Ramsey-Musolf, H. Patel, S. Petcov, P. Schwaller, S. Tulin and P. Ullio for helpful advice and suggestions.  W.-C.~H. would like to thank the hospitality of Physics Division of National Center for Theoretical Sciences (NCTS) in Taiwan, where part of this work was performed.  This work is supported in part under US DOE contract No. DE-FG02-91ER40684.

\section{Appendix}
Here, we discuss how we can use additional multiplets to loosen the restrictions placed on the multiplet quantum numbers by the requirement of anomaly cancellation.  We start with adding one more copy (or ``family'') of $\chi_{n_i,Y_i}$ fields, i.e.,
\begin{equation}
\label{eq:qnumsapx}
\chi_{n',Y'}, \chi_{n',Y'-1}, \chi_{n'+1,\frac{1}{2}-Y'}, \chi_{n'-1,\frac{1}{2}-Y'+A'}
\end{equation}
 Following the same procedure as those that led to Eqs. (\ref{eq:anomy}) gives the analogous expressions
\begin{eqnarray}
&&(n-1) A+(n'-1) A'=0 \nonumber\\
&&3(n-1)A (\frac{1}{2}-Y)^2+3(-\frac{n}{2}+A^2 (n-1)) (\frac{1}{2}-Y)+(n-1)A^3\nonumber\\&+&3(n'-1)A' (\frac{1}{2}-Y')+3(-\frac{n'}{2}+A'^2 (n'-1)) (\frac{1}{2}-Y')+(n'-1)A'^3=0 \label{eq:anomyapx}  \\
&&(\frac{n}{2})(\frac{1}{2}-Y)+f(n-1)A + (\frac{n'}{2})(\frac{1}{2}-Y')+f(n'-1)A'=0,\nonumber
\end{eqnarray}
where again $A$ and $A'$ can take on the values $-1, 0,$ and $1$.  

We now seek solutions to these equations such that $Y,Y'\ne \frac{1}{2}$ to avoid having multiplets with zero hypercharge and to avoid having two multiplets with equal $n_i$ but opposite $Y_i$, like we saw in the single-family scenario.  Let us examine the first of these three equations.  The solutions to this equation lead us to consider three cases.
\subsection{Case 1:  $n=n'$}
If $n=n'$, from the first of Eqs. (\ref{eq:anomyapx}), $n=n'=1$ or $A=-A'$. Then, the second and third equations  yield $(\frac{1}{2}-Y)=-(\frac{1}{2}-Y')$.  This, however, allows a mass term  $ \chi_{n+1,\frac{1}{2}-Y}\chi_{n'+1,\frac{1}{2}-Y'}$, so we discard this option.
\subsection{Case 2:  $n\ne n'$, $n,n'\ne 1$}
If $n\neq 1$, $n' \neq 1$ and $n \neq n'$, from the first equation of (\ref{eq:anomyapx}), $A=A'=0$. Then, the second and third equation of (\ref{eq:anomyapx}) both are fulfilled if $(\frac{1}{2}-Y)=-\frac{n'}{n}(\frac{1}{2}-Y')$.
\subsection{Case 3:  $n=1$, $n\ne n'$}
For $n=1$, $n\ne n'$, the first of Eqs. (\ref{eq:anomyapx}) gives $A'=0$.  Then, the second and third equations are satisfied if $(\frac{1}{2}-Y)=-n' (\frac{1}{2}-Y')$.

Of course, there is also a case analogous to Case 3 but with the two families interchanged.

Additionally, we note that, for integer charges, even $n$ implies half-integer $Y$, while odd $n$ implies integer $Y$; similar relations hold for $n'$ and $Y'$.  For Case 2, this implies that $n$ and $n'$ must either both be even or both be odd.  For case 3, it implies that $n'$ is odd.

Of course, more general scenarios are possible; as an example, the SM accomplishes anomaly cancellation for a given family using one $n=1$ lepton polyplet and three (colored) $n=1$ quark polyplets.

\bibliographystyle{h-physrev}
%\bibliography{topbib}

\end{document}